\documentclass[prd,preprintnumbers,showkeys,nofootinbib,11pt]{article}
\usepackage{verbatim}
\usepackage{graphicx,amsmath,amssymb,epsfig}
\usepackage[active]{srcltx}
\def\beq{\begin{equation}}
\def\eeq{\end{equation}}

\def\beqn{\begin{eqnarray}}
\def\eeqn{\end{eqnarray}}

\newcommand{\Rmnum}[1]{\expandafter\@slowromancap\romannumeral #1@}

\begin{document}

\title{\large \bf  BFKL approach  and   $2 \to 5$  MHV amplitude }
\author{\large J.~Bartels$^{1}$,  A.~Kormilitzin$^{2,3}$,  L.~N. Lipatov$^{1,4}$ and
A.~Prygarin$^{5}$ \bigskip \\
{\it
$^1$~II. Institute of  Theoretical Physics, Hamburg University, Germany}
\\
{\it  $^2$~Tel-Aviv University, Ramat Aviv, Israel} 
\\
{\it  $^3$~Physics Department, Ben-Gurion University of the Negev, Beer Sheva 84105, Israel} 
\\
{\it  $^4$~St. Petersburg Nuclear Physics Institute, Russia} 
\\
{\it  $^5$~Brown University, Providence, RI, USA}}

\maketitle

\vspace{-9cm}
\begin{flushright}
%$~~$\\
%DESY-XX-XXX
\end{flushright}
\vspace{8cm}

\abstract{
We study MHV amplitude for the $2 \to 5$ scattering in the multi-Regge kinematics. The Mandelstam cut correction to the BDS amplitude is calculated in the leading logarithmic approximation~(LLA) and the  corresponding  remainder function is given to any loop order in a closed integral form. We show that the   LLA remainder function at two loops for $2 \to 5 $ amplitude can be written as a sum of two $2 \to 4 $  remainder functions due to  recursive properties of the leading order impact factors. We also make some generalizations for the MHV amplitudes with more external particles. The results of the  present study are in  agreement  with     all leg two loop symbol derived by Caron-Huot as shown in a parallel paper of one of the authors with collaborators.

}

\newpage

\section{Introduction}

We further study the high energy behavior of the scattering amplitudes in $\mathcal{N}=4$ supersymmetric Yang Mills~(SYM) theory. The  recent impressive advances in the field of the  Maximally Helicity Violating~(MHV) amplitudes motivated us to continue the programm  of applying  the  Balitsky-Fadin-Kuraev-Lipatov~(BFKL)~\cite{BFKL} approach to the  MHV amplitudes, initiated by two of the authors in collaboration with A.~Sabio~Vera in Ref.~\cite{BLS1,BLS2}. In Ref.~\cite{BLS1} the   analytic properties of the Bern-Dixon-Smirnov~(BDS) ansatz~\cite{BDS} were tested    and
 found to contradict the well known  high energy behavior of the multileg scattering amplitudes starting at  two loops for  six  external particles. A similar  conclusion was drawn  from  the strong
coupling side by Alday and Maldacena~\cite{Alday:2007he}.
The analytic properties of  the BDS amplitude were also studied in Refs.~\cite{Schnitzer1,Schnitzer2,DelDuca:2008jg}.

Two of the authors with a collaborator in Ref.~\cite{BLS2} derived a closed integral form of the all loop correction to the $n=6$ BDS formula in the leading logarithmic approximation~(LLA) and calculated it analytically at two loops.
The need  of correcting the BDS amplitude by so-called remainder function triggered a big activity in this direction, which resulted in introducing novel and powerful  computation techniques. The most recent one is the use of the symbol  successfully applied by Goncharov, Spradlin, Vergu and Volovich~\cite{GSVV} to calculating  the six-gluon two-loop remainder function from Wilson loops with light-like edges~\cite{hexagon,DDS}. Their result was shown by two of the authors~\cite{LP1} to reproduce the BFKL predictions in the multi-Regge kinematics and was used to calculate  the  next-to-leading impact factor~\cite{LP2} needed in the BFKL approach.  The three loop BFKL predictions by three of the authors~\cite{LP2,BLP3to3,Lipatov:2010qf,BLPreview} were used among other things  in writing the symbol of the remainder function in general kinematics by Dixon, Henn and Drummond~\cite{Dixon3}.  Their findings are in agreement with next-to-leading three loop calculations of one of the authors in collaboration with V.~Fadin~\cite{Fadin:2011we}.

The Regge factorization made it possible to write the  all loop  LLA remainder function as a closed  integral in Ref.~\cite{BLS2} and calculate it analytically to all loops in a more restrictive kinematics~(multi-Regge and collinear) by three of the authors in Ref.~\cite{BLPope}. This   was shown  to be consistent with the OPE expansion formula for the remainder function in the collinear limit derived by
 Alday,  Gaiotto,  Maldacena, Sever and  Vieira~\cite{Alday:2010ku}. An intriguing similarity of  the OPE formulae~\cite{Sever1,Sever2,Sever3,Sever4} and the BFKL expressions suggest some deep relation between these two approaches that describe very  different kinematic regimes. These relation seems to be analogous to the relation between Dokshitzer-Gribov-Lipatov-Altarelli-Parisi~(DGLAP)~\cite{DGLAP} and BFKL equations.

The extensive use of symbol in the recent studies of the MHV and $\text{N}^k$MHV amplitudes made it possible to derive important analytic results  in a series of very interesting publications~\cite{CaronHuot:2011kk,Dixon:2011nj,Duhr:2011zq,CaronHuot2}.
The symbol for all leg two loop MHV amplitude derived by Caron-Huot in Ref.~\cite{CaronHuot2} using an extended superspace is of particular interest to us because it is directly related to the main results of the present paper as follows.

In the next section we consider the $2 \to 5$ MHV scattering amplitude in a kinematic region, where Mandelstam cut gives a non-vanishing contribution. We call this region the Mandelstam region. Using the BFKL approach we derive the LLA remainder function  in closed integral form of eq.~(\ref{R25L}) and calculate it analytically at two loops given by eq.~(\ref{R25compact}). It turns out that the $2 \to 5$ remainder function can be written as a sum  of two $2 \to 4$ remainder functions due to the recursive properties of the leading order impact factors in the BFKL approach depicted in Fig.~\ref{fig:F23tildeRed}.
We generalize this result for the case of the $2 \to 2+(n-4)$ scattering and write the corresponding two loop  LLA answer for an arbitrary $n$ as a sum of six-particle remainder functions in eq.~(\ref{R2n4}) and eq.~(\ref{R2nmk}) in  a particular Mandelstam region. We choose this  region such that  we do not have contributions of  Bartels-Kwiecinski-Praszalowicz~(BKP)~\cite{Bartels:1980pe,Kwiecinski:1980wb} states that    appear  in other regions  for  $n \geq 8$.  The Mandelstam regions with BKP states can be identified using dispersion representation of the $2 \to 2+(n-4)$ similar to one in Ref.~\cite{BLS1}. The number of terms in the dispersion representation is determined by the Steinmann relations~\cite{Steinmann}.

The remainder functions in eq.~(\ref{R25compact}), eq.~(\ref{R2n4}) and  eq.~(\ref{R2nmk}) are the main result of the present paper.
 The parallel study of one of the authors in collaboration with Spradlin, Vergu and Volovich~\cite{PSVV} shows agreement between the BFKL calculations of  the current  paper and  the  symbol derived by Caron-Huot in Ref.~\cite{CaronHuot2} for the   two loop remainder function with an arbitrary number of external particles.

Some relevant calculations are presented in the Appendices.

\section{$2 \to 5$  amplitude}
\begin{figure}[h]
  \begin{center}
    %\showthe\columnwidth % Use this to determine the width of the figure.
    \includegraphics[width=0.3\columnwidth]{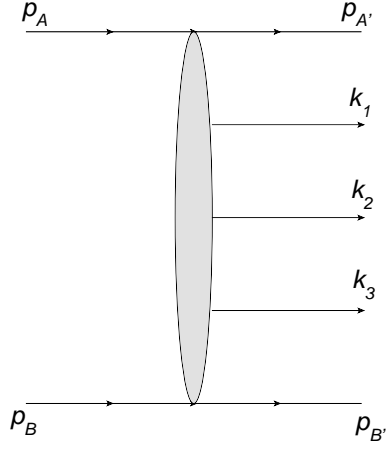}
    \caption{ The $2 \to 5 $ amplitude.  The produced particles $k_i$ are strongly ordered in   rapidity in the multi-Regge kinematics.  } \label{fig:multi25N}
  \end{center}
\end{figure}

In this section we
study the high energy behavior of the  $2 \to 5 $  scattering amplitude shown in Fig.~\ref{fig:multi25N} in $\mathcal{N}=4$ supersymmetric Yang Mills~(SYM) theory.     In the leading logarithmic approximation~(LLA) only gluons contribute since in this limit  $t$-channel exchanges  are dominated by particles with highest spin. This fact allows us to apply the QCD-based BFKL approach to calculating the corrections to the BDS amplitude with the leading logarithmic accuracy.

The Mandelstam invariants for the $2 \to 5$ amplitude are given by
\beqn
&& s=(p_A+p_B)^2, \; s_{A'123}=(p_{A'}+k_1+k_2 +k_3)^2, \; s_{B'123}=(p_{B'}+k_1+k_2 +k_3)^2,
 \\
 &&\; s_{123}=(k_1+k_2 +k_3)^2, \; s_{A'12}=(p_{A'}+k_1+k_2 )^2, \; s_{B'23}=(p_{B'}+k_2 +k_3)^2, \; s_{1}=(p_{A'}+k_1)^2, \nonumber
  \\
  &&
   s_{2}=(k_1+k_2)^2, \; s_{3}=(k_2 +k_3)^2, \;s_{4}=(p_{B'} +k_3)^2, \; t_{1}=(p_{A'}-p_A)^2, \;
   t_{2}=(p_{A'}+k_1-p_A)^2,
   \nonumber\\
   &&
    \; t_{3}=(p_{A'}+k_1+k_2-p_A)^2, \; t_{4}=(p_{A'}+k_1+k_2+k_3-p_A)^2=(p_{B'}-p_B)^2. \nonumber
\eeqn

The multi-Regge kinematics is characterized by a strong ordering in the rapidity of the produced particles  \beqn\label{multiRegge}
s \gg  s_{A'123},  s_{B'123} \gg  s_{123}, s_{A'12}, s_{B'23} \gg s_{1}, s_{2}, s_{3}, s_{4} \gg -t_1, -t_2, -t_3, -t_4.
\eeqn

It is useful to introduce Sudakov parametrization
\beqn\label{Sudakov}
k_{i}^{\mu}=\alpha_i p^\mu_{A}  +\beta_i p^\mu_{B} +k^{\mu}_{i\perp}
\eeqn
for which the multi-Regge kinematics in eq.~(\ref{multiRegge}) reads
\beqn\label{multiReggeA}
1 \gg \alpha_1  \gg \alpha_2 \gg \alpha_3 >0.
\eeqn
The $\beta_i$ components of $k_{i}^{\mu}$ are also ordered due to the on-shellness  condition
\beqn
\beta_i=\frac{-k^2_{i\perp}}{ \alpha_i s }=\frac{\mathbf{k}^2_i}{ \alpha_i s },
\eeqn
where $k^2_{i\perp}=-\mathbf{k}^2_i$. In the rest of the paper we use bold script to denote the transverse momenta.

The  high energy scattering amplitudes have contributions of the Regge poles (poles in the complex angular momentum plane)  and Mandelstam~(Regge) cuts. In planar amplitudes the contributions of the Mandelstam cuts cancel out in most regions while in some regions~(Mandelstam regions) they give logarithmically    divergent terms~(see Ref.~\cite{BLS1} for more details).

\begin{figure}[h]
  \begin{center}
    %\showthe\columnwidth % Use this to determine the width of the figure.
    \includegraphics[width=0.3\columnwidth]{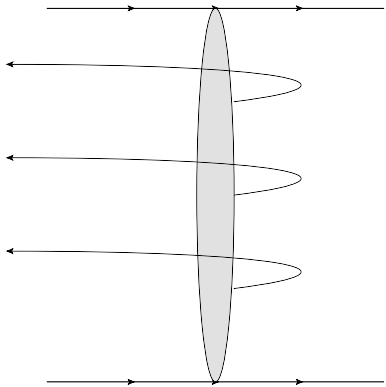}
    \caption{ Mandelstam region of the $2 \to 5$ amplitude with a non-vanishing contribution  of the discontinuities  in   $s$ and $s_{123}$ to the remainder function.  }\label{fig:25flipped3}
  \end{center}
\end{figure}

In this section we consider one of the Mandelstam regions for  the  $2 \to 5$ scattering amplitude shown in Fig.~\ref{fig:25flipped3}, where we flip all three produced particles and have a non-vanishing Mandelstam cut contribution  that corresponds to the discontinuity  in $s$ and $s_{123}$. In this region
\beqn\label{MandRegion}
s, s_{2}, s_{3}, s_{123} >0
\eeqn
and all other energy invariants are negative.

 For the scattering amplitudes with $n$ external particles being on-shell we have $3n-15$ independent cross ratios
\beqn
u=\frac{x^2_{ij} x^2_{kl}}{x^2_{kj} x^2_{il}}
\eeqn
written in terms of the dual coordinates defined by
\beqn
p_{i}=x_{i}-x_{i+1}.
\eeqn
In the case of the $2 \to 5$ amplitude we have $n=7$ and thus $6$ independent cross ratios. However, it is beneficial for our purposes to follow the notation of the BDS paper and use $7$ cross ratio, when we parameterize them in terms of the physical momenta.
In the multi-Regge kinematics eq.~(\ref{multiReggeA}) the cross ratios we use   are given by
\beqn\label{Uij}
 &&
 u_{73}=\frac{x^2_{13}x^2_{47}}{x^2_{14}x^2_{73}}=\frac{(-s_2)(-s_{A'123})}{(-s_{123})(- s_{A'12})}\simeq 1,
 \\
&&
 u_{14}=\frac{x^2_{51}x^2_{24}}{x^2_{14}x^2_{25}}=\frac{(-s_{123B'})(- s_3)}{(-s_{123})(- s_{23B'})} \simeq 1, \nonumber
 \\
&&
 u_{25}=\frac{x^2_{62}x^2_{35}}{x^2_{25}x^2_{36}}=\frac{(-s_4)(- t_2)}{(-s_{23B'})(- t_3)}\simeq \frac{\mathbf{q}_2^2 \alpha_3}{\mathbf{q}_3^2 \alpha_2} \simeq 0, \nonumber \\
&&
 u_{36}=\frac{x^2_{73}x^2_{46}}{x^2_{36}x^2_{47}} =\frac{(- s_{A'12})(- t_4)}{ (-s_{A'123})(-t_3)}\simeq \frac{\mathbf{q}_4^2 \mathbf{k}_2^2 \alpha_3}{\mathbf{q}_3^2  \mathbf{k}_3^2\alpha_2}  \simeq 0, \nonumber
 \\
&&
 u_{47}=\frac{x^2_{14}x^2_{57}}{x^2_{47}x^2_{51}}=\frac{(-s_{123})(- s)}{(-s_{123B'})(- s_{A'123})} \simeq 1,
 \nonumber
 \\
&&
 u_{51}=\frac{x^2_{25}x^2_{61}}{x^2_{51}x^2_{62}}=\frac{(-t_1)(- s_{23B'})}{ (-t_2)(- s_{123B'})}\simeq \frac{\mathbf{q}_1^2 \alpha_2}{\mathbf{q}_2^2 \alpha_1}
 \simeq 0, \nonumber
 \\
&&
 u_{62}=\frac{x^2_{63}x^2_{72}}{x^2_{62}x^2_{73}}=\frac{(-s_1)(- t_3)}{(-s_{A'12})(- t_2)}\simeq \frac{\mathbf{q}_3^2 \mathbf{k}_1^2 \alpha_2}{\mathbf{q}_2^2
  \mathbf{k}_2^2 \alpha_1}  \simeq 0. \nonumber
\eeqn

In this kinematics we consider a symmetric subregion, where we have  $\alpha_2/\alpha_1 \simeq \alpha_3/\alpha_2=\delta$ for $\delta$ being a small parameter.
Note that three of them go to  unity with  different rates
\beqn\label{1minusU}
&&
1-u_{14} \simeq \frac{(\mathbf{k}_2 +\mathbf{k}_3)^2}{s_3} \simeq \frac{(\mathbf{k}_2+\mathbf{k}_3)^2}{\mathbf{k}_3^2} \frac{\alpha_3}{\alpha_2} \propto \delta,
\\
&&
1-u_{73} \simeq \frac{(\mathbf{k}_1 +\mathbf{k}_2)^2}{s_2} \simeq \frac{(\mathbf{k}_1+\mathbf{k}_2)^2}{\mathbf{k}_2^2} \frac{\alpha_2}{\alpha_1} \propto \delta, \nonumber
\\
&&
1-u_{47} \simeq \frac{(\mathbf{k}_1+\mathbf{k}_2 +\mathbf{k}_3)^2}{s_{123}} \simeq \frac{(\mathbf{k}_1+\mathbf{k}_2+\mathbf{k}_3)^2}{\mathbf{k}_3^2} \frac{\alpha_3}{\alpha_1} \propto \delta^2.
\nonumber
\eeqn

In the Mandelstam region of eq.~(\ref{MandRegion}) shown in Fig.~\ref{fig:25flipped3} only one cross ratio has a non-zero phase
\beqn
u_{47}=|u_{47}| \;e^{-i2\pi}.
\eeqn
It is worth emphasizing that this is also the only cross ratio that depends on the transverse mass $(\mathbf{k}_1+\mathbf{k}_2+\mathbf{k}_3)^2$ of the bunch of the produced particles that are flipped in this Mandelstam region. It is also the most rapidly approaching the unity cross ratio as one can see from eq.~(\ref{1minusU}). We use these two facts later  to identify the cross ratio that has the phase for similar Mandelstam regions in the scattering amplitudes with a larger number of external gluons.

\begin{figure}[h]
  \begin{center}
    %\showthe\columnwidth % Use this to determine the width of the figure.
    \includegraphics[width=0.35\columnwidth]{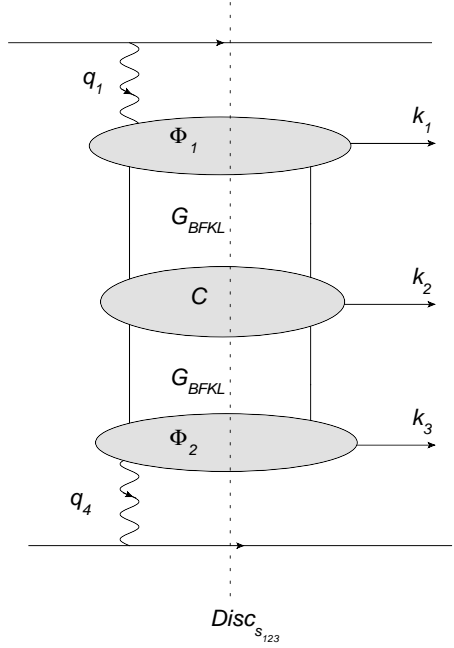}
    \caption{  The $2 \to 5$ amplitude factorized into blocks. $\Phi_i$ represent impact factors, $G_{BFKL}$  illustrates    propagation of the  BFKL state and $C$ denotes the central emission block. The dashed line stands for the discontinuity in $s_{123}$.  } \label{fig:impact7}
  \end{center}
\end{figure}

Now we apply the BFKL approach to calculating   the discontinuity in $s_{123}$ of the $2 \to 5 $ scattering amplitude to the leading logarithmic accuracy.
In a way analogous to the $2 \to 4 $ amplitude considered in Ref.~\cite{BLS2} we can take advantage of the Regge factorization, which allows us to decompose the amplitude into several blocks that we calculate separately. These blocks are shown in Fig.~\ref{fig:impact7}.
The impact factors $\Phi_i$ are the same as in the  case of the  $ 2 \to 4$ amplitude. The  BFKL Green function   $G_{BFKL}$  is universal and  corresponds to the homogeneous octet-BFKL equation for the wave function $f$ with removed propagators
\beqn
E f = \tilde{H} f.
\eeqn
The detailed discussion of this equation is presented in Ref.~\cite{BLS2} and here we only need its eigenvalues
\beqn\label{Enun}
E_{\nu, n} =-\frac{1}{2} \frac{|n|}{\nu^2+\frac{n^2}{4}}
+\psi \left(1+i\nu +\frac{|n|}{2} \right)
+\psi \left(1-i\nu +\frac{|n|}{2} \right)
-2 \psi(1)
\eeqn
 and eigenfunctions
 \beqn
 f_{\nu,n} (k,q)=\left(\frac{k}{q-k}\right)^{i \nu +\frac{n}{2}} \left(\frac{k^*}{q^*-k^*}\right)^{i \nu +\frac{n}{2}}
 \eeqn
 with their completeness  condition
 \beqn\label{compBFKL}
 \sum_{n=-\infty}^{\infty} \int_{-\infty}^{+\infty} d \nu f^*_{\nu,n} (k',q') f_{\nu,n} (k,q)
 = 2 \pi^2 \delta^2 (k'-k) \frac{|k|^2 |q-k|^2}{|q|^2}.
 \eeqn

\begin{figure}[h]
  \begin{center}
    %\showthe\columnwidth % Use this to determine the width of the figure.
    \includegraphics[width=0.3\columnwidth]{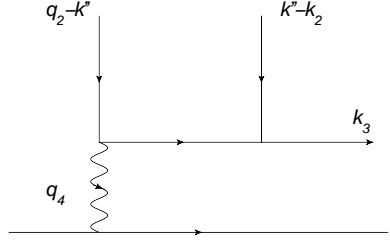}
    \caption{ The impact factor $\Phi_2$ in Fig.~\ref{fig:impact7}.   }\label{fig:F2}
  \end{center}
\end{figure}

The impact factor $\Phi_2$ in Fig.~\ref{fig:F2}
  \beqn\label{Phi2}
  \Phi_2=\frac{k_3 (k''-q_2)}{q_3 (k''-k_2-k_3)}
  \eeqn

  was calculated  in the terms of the anomalous dimension $\nu$ and the conformal spin $n$ in Ref.~\cite{BLS2} (note a redefinition of momenta)
\beqn\label{chi2}
&&
\chi_2= \int \frac{d^2 k''}{2 \pi }
\frac{|q_3|^2}{ |k''-k_2|^2 |q_2 -k''|^2}
\left(\frac{q_2-k''}{k''-k_2}\right)^{i\mu +\frac{m}{2}}
\left(\frac{{q_2}^*-{k''}^*}{{k''}^*-{k_2}^*}\right)^{i\mu -\frac{m}{2}}
\frac{k_3 (k''-q_2)}{q_3 (k''-k_2-k_3)}\;\;\; \\
&&
\;\;\;\;\;\;\;\;\;\;\;
=-\frac{1}{2}\frac{1}{i\mu-\frac{m}{2}}\left(\frac{q^*_4}{k^*_3}\right)^{i\mu -m/2}
 \left(\frac{q_4}{k_3}\right)^{i\mu +m/2}. \;\;\;\nonumber
\end{eqnarray}
In a similar way one finds
\beqn\label{chi1}
&& \chi_1=\frac{1}{2}\frac{1}{i\nu+\frac{n}{2}}\left(-\frac{q_1}{k_1}\right)^{-i\nu -n/2}
 \left(-\frac{q^*_1}{k^*_1}\right)^{-i\nu +n/2}
 =
 \frac{1}{2}\frac{(-1)^n}{i\nu+\frac{n}{2}}\left(\frac{q_1}{k_1}\right)^{-i\nu -n/2}
 \left(\frac{q^*_1}{k^*_1}\right)^{-i\nu +n/2}.  \;\;\;
\eeqn

\begin{figure}[h]
  \begin{center}
    %\showthe\columnwidth % Use this to determine the width of the figure.
    \includegraphics[width=0.4\columnwidth]{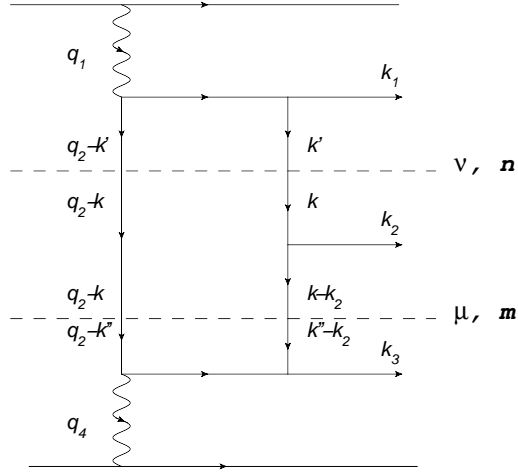}
    \caption{ Decomposition of $2 \to 5$ amplitude into two impact factors and a central emission block. The dashed lines denote   insertions of the BFKL eigenvalue.  } \label{fig:7dashBKPL}
  \end{center}
\end{figure}

A new piece in the $2 \to 5$ amplitude compared to the $2 \to 4$ amplitude is the central emission block $C$ in Fig.~\ref{fig:impact7}.  In order to find the central emission block $C$ we insert the completeness condition of the BFKL eigenfunctions  between each two adjacent produced particles $k_i$ as shown in Fig.~\ref{fig:7dashBKPL}. Each such insertion is denoted by a dashed line that also denotes   insertions of the BFKL eigenvalue of eq.~(\ref{Enun}) when we go to higher loops.

\begin{figure}[h]
  \begin{center}
    %\showthe\columnwidth % Use this to determine the width of the figure.
    \includegraphics[width=0.4\columnwidth]{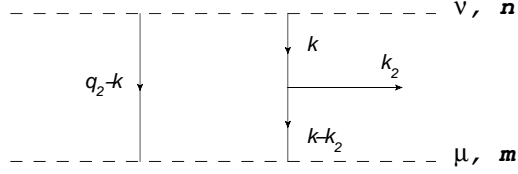}
    \caption{ The central emission block $C$ in Fig.~\ref{fig:7dashBKPL}. }\label{fig:centerBKPL}
  \end{center}
\end{figure}

In the Appendix~\ref{sec:block} we calculate   the central emission block $C$ illustrated in Fig.~\ref{fig:centerBKPL} in terms of the conformal spins $n$ and $m$ as well as the anomalous dimensions $\nu$ and $\mu$
and the result reads
\beqn\label{C}
 C= %\frac{1}{|q_2|^2} \frac{q_2 (q_2^*-k_2^*)}{k_2^*} \frac{1}{|q_2-k_2|^2}     \;\;
-\frac{1}{2}\left(\frac{q_3^*}{k_2^*}\right)^{i\nu-\frac{n}{2}}\left(\frac{q_3}{k_2}\right)^{i\nu+\frac{n}{2}}
\left(\frac{q_2}{k_2}\right)^{-i\mu-\frac{m}{2}}\left(\frac{q_2^*}{k_2^*}\right)^{-i\mu+\frac{m}{2}} \;\; B,
\eeqn
where $I$ is given by
\beqn
B=
(-1)^{m-1}\frac{\Gamma (1-i\nu -n/2) }{\Gamma (1+i\nu -n/2)}\,
\frac{\Gamma (+i\mu +m/2) }{\Gamma (-i\mu +m/2)}\,\frac{\Gamma (i(\nu -\mu ) +(m-n)/2) }{\Gamma (1-i(\nu -\mu ) +(m-n)/2)}.
\eeqn

It is worth emphasizing that $C$ in eq.~(\ref{C}) reduces to the form of the impact factors in eq.~(\ref{chi2}) and eq.~(\ref{chi1}) in the following limits
\beqn\label{limits}
&&
\lim_{\mu \to 0 } C_{m=0}=-\frac{1}{2}\frac{1}{i\nu-\frac{n}{2}}\left(\frac{q^*_3}{k^*_2}\right)^{i\nu -n/2}
 \left(\frac{q_3}{k_2}\right)^{i\nu +n/2}, \;\; \\
 &&
 \lim_{\nu \to 0 } C_{n=0}=\frac{1}{2}\frac{(-1)^m}{i\mu+\frac{n}{2}}\left(\frac{q^*_2}{k^*_2}\right)^{-i\mu +m/2}
 \left(\frac{q_2}{k_2}\right)^{-i\mu +m/2}. \nonumber
\eeqn

After calculating all building blocks for the discontinuity in $s_{123}$ of the $2 \to 5 $ scattering amplitude in Fig.~\ref{fig:impact7} we  convolve impact factors $\Phi_1$, $\Phi_2$ and the central emission block $C$ together with the BFKL eigenvalues $E_{\nu,n}$ and $E_{\mu,m}$ and obtain
\beqn\label{disc}
 \mathcal {\Im}_{s_{123}}  \mathcal{M}_{2 \to 5} = \frac{a}{ \pi}
 s_2^{\omega (t_2)} s_3^{\omega (t_3)}\;\;
  \text{Reg}_{s_{123}}\;\;
 \sum_{n=-\infty}^{+\infty} \sum_{m=-\infty}^{+\infty} \int_{-\infty}^{+\infty} d \nu \int_{-\infty}^{+\infty} d \mu
 \;\;
\chi_1 \;\; C \;\; \chi_2 \left(\frac{s_{123}}{s_0}\right)^{-a  (E_{\nu,n}+E_{\mu,m})}.
\eeqn
The expression in eq.~(\ref{disc}) is divergent at one loop and should be regularized. We fix regularization in Appendix~\ref{sec:twoloops} by comparing  eq.~(\ref{disc})  to  BDS amplitude at one loop. At higher loops this integral is finite due to the fact that $E_{\nu, n} \to 0$ for $n=0$ and $\nu \to 0$ and all infrared divergencies are included in the gluon Regge trajectories   $\omega (t_i)$  given by
\beqn
\omega(t)=a \left(  \frac{1}{\epsilon} -\ln \frac{(-t)}{\mu^2} \right),
\eeqn
where $d=4-2 \epsilon$ and
\beqn
a=\frac{\alpha_s N_c}{2 \pi} \left(4 \pi e^{-\gamma}\right)^{\epsilon}.
\eeqn

\begin{figure}[h]
  \begin{center}
    %\showthe\columnwidth % Use this to determine the width of the figure.
    \includegraphics[width=0.35\columnwidth]{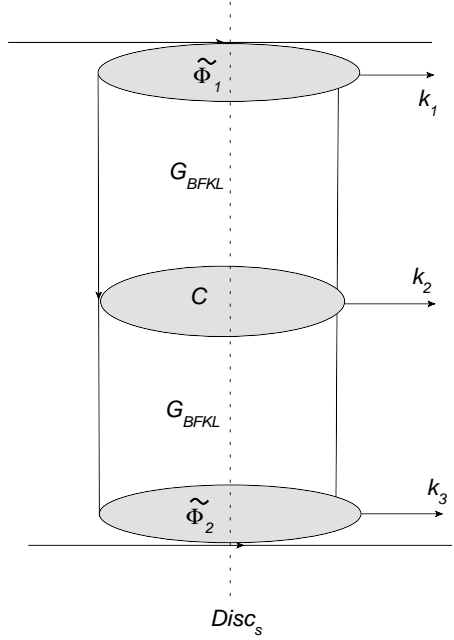}
    \caption{ Decomposition of the $2 \to 5$ amplitude for the discontinuity in $s$.  The amplitude is built of  two  impact factors $\tilde{\Phi}_i$, central emission block   $C$ and the  BFKL Green function  $G_{BFKL}$.  }
    \label{fig:impact7DiscS}
  \end{center}
\end{figure}

 Then the Mandelstam  cut correction to the BDS amplitude at the leading logarithmic accuracy reads
 \beqn\label{MBDS}
 M=M^{BDS} \left(1+ \frac{2 i  a}{\pi}\sum_{n=-\infty}^{+\infty} \sum_{m=-\infty}^{+\infty} \int_{-\infty}^{+\infty} d \nu \int_{-\infty}^{+\infty} d \mu
 \;\;
\chi_1 \;\; C \;\; \chi_2 \left( \left( \frac{s_{123}}{s_0}\right)^{-a  (E_{\nu,n}+E_{\mu,m})}-1\right)\right).
 \eeqn
The minus unity in the brackets on the right hand side of eq.~(\ref{MBDS}) removes the divergent one loop contribution.
The choice of the energy scale $s_0$ does not change the result in  the leading logarithmic  approximation and thus it is dictated only by the requirement of the dual conformal invariance and the Regge factorization.
Among other possible choices we prefer
\beqn\label{s123k}
\frac{s_{123}}{s_{0}} = \frac{1}{\sqrt{u_{25} \; u_{36} \; u_{51} \; u_{62}}} \simeq \frac{\alpha_1}{\alpha_3}  \frac{|\mathbf{q}_2|\; |\mathbf{q}_3| \; |\mathbf{k}_3|} {|\mathbf{q}_1| \; |\mathbf{q}_4| \; |\mathbf{k}_1|},
\eeqn
where  the cross ratios $u_{ij}$ are given in  eq.~(\ref{Uij}). So that the remainder function $R$ defined by
\beqn
M=M^{BDS} R_{2 \to 5}, \;\;\; R_{2 \to 5}= 1+a^2 R_{2 \to 5}^{(2)} +a^3 R_{2 \to 5}^{(3)}+...
\eeqn
reads
\beqn\label{R25L}
R_{2 \to 5}^{(\ell)} =\frac{i 2 }{ \pi } \frac{(-1)^{\ell-1}}{(\ell-1)!} \ln^{\ell-1} \left( \frac{1}{\sqrt{u_{25} \; u_{36} \; u_{51} \; u_{62}}} \right)
 \sum_{n=-\infty}^{+\infty} \sum_{m=-\infty}^{+\infty} \int_{-\infty}^{+\infty} d \nu \int_{-\infty}^{+\infty} d \mu
 \;\;
\chi_1 \;\; C \;\; \chi_2  \;\; (E_{\nu,n}+E_{\mu,m})^{\ell-1}
\eeqn
where the product  $\chi_1 \;C\; \chi_2$~( see   eqs.~(\ref{chi2})-(\ref{C})) can be written as
\beqn\label{chi1Cchi2}
 \chi_1 \;C\; \chi_2 =\frac{(-1)^{n+m}}{8}
\frac{\Gamma (-i\nu -\frac{n}{2}) }{\Gamma (1+i\nu -\frac{n}{2})}\,
\frac{\Gamma (i\mu +\frac{m}{2}) }{\Gamma (1-i\mu +\frac{m}{2})}\,\frac{\Gamma (i(\nu -\mu ) +\frac{m-n}{2}) }{\Gamma (1-i(\nu -\mu ) +\frac{m-n}{2})}
 w_1^{i\nu+\frac{n}{2}} (w^*_1)^{i\nu-\frac{n}{2}}
 w_2^{i\mu+\frac{m}{2}} (w^*_2)^{i\mu-\frac{m}{2}}.
\eeqn
 Here we introduce a complex variables $w_i$ expressed in terms of the complex  transverse momenta
 \beqn\label{anharmonic}
 w_1=\frac{k_1 q_3 }{  q_1 k_2}, \;\; w_2=\frac{k_2 q_4 }{  q_2 k_3}.
 \eeqn
The variables $w_i$ are cross ratios in the dual space of the transverse momentum.
To this point we considered only the discontinuity in $s_{123}$. However  there is also discontinuity in $s$ for the $2 \to 5$ scattering amplitude in the Mandelstam region shown in Fig.~\ref{fig:25flipped3}.  The discontinuity in $s$ of the $2 \to 5$  scattering amplitude is illustrated in Fig.~\ref{fig:impact7DiscS}.

\begin{figure}[h]
  \begin{center}
    %\showthe\columnwidth % Use this to determine the width of the figure.
    \includegraphics[width=0.35\columnwidth]{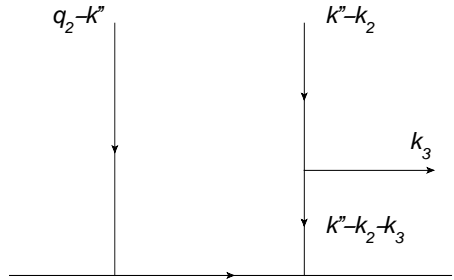}
    \caption{ The lower impact factor $\tilde{\Phi}_2$ in Fig.~\ref{fig:impact7DiscS}.  }
    \label{fig:F2tilde}
  \end{center}
\end{figure}

The only difference  between calculation of the discontinuity in $s$ and the discontinuity in $s_{123}$ is that the impact factors $\Phi_i$ should be replaced by a slightly different impact factors $\tilde{\Phi}_{i}$. For example, for $i=2$ the impact factor  for the discontinuity in $s$  is  depicted  in  Fig.~\ref{fig:F2tilde}.

The impact factor $\tilde{\Phi}_2$ was calculated in Ref.~\cite{BLS2}
\beqn\label{Phi2tilde}
 \tilde{ \Phi_2}=\frac{q_4 (k''-k_2)}{q_3 (k''-k_2-k_3)}.
  \eeqn
  Due to the fact that
\beqn\label{PhiPhitilde}
\Phi_2+\tilde{ \Phi_2}=1
\eeqn
the discontinuity in $s$ is expressed through  integral we had before for the discontinuity in $s_{123}$, though  with a different regularization at one loop in full analog with the $2 \to 4$ case. Thus starting at two loops the contribution  of the discontinuity in $s$ to the remainder function is the same as of the discontinuity in $s_{123}$.

The LLA remainder function $R_{2 \to 5 }$ in eq.~(\ref{R25L}) at two loops to the leading logarithmic accuracy  was calculated in Appendix~\ref{sec:twoloops}
\beqn\label{R25L2}
&&
R^{(2)}_{2 \to 5}= \frac{i \pi }{ 2} \ln \sqrt{u_{25} \; u_{36} \; u_{51} \; u_{62}}
\left( \ln |1+w_2 +w_1 w_2|^2 \ln \left|\frac{1+w_2+w_1 w_2 }{w_2 (1+w_1)}\right|^2 \right.
\\
&&
\left.
  \hspace{2cm}+ \ln \left|\frac{1+w_2+w_1 w_2}{w_1 w_2}\right|^2
\ln \left|\frac{1+w_2+w_1 w_2}{1+w_2}\right|^2  \right) \nonumber
\\
&&
\simeq -\frac{i \pi}{2} \ln \left(\frac{s_{123}}{s_0}\right)  \left(
\ln \frac{\mathbf{q}_3^2 (\mathbf{k}_1 +\mathbf{k}_2+\mathbf{k}_3)^2}{\mathbf{q}_1^2\mathbf{k}_3^2}
\ln \frac{\mathbf{q}_3^2 (\mathbf{k}_1 +\mathbf{k}_2+\mathbf{k}_3)^2}{\mathbf{q}_4^2(\mathbf{k}_1 +\mathbf{k}_2)^2}
\right. \nonumber
\\
&&
\hspace{2cm}
\left.
+
\ln \frac{\mathbf{q}_2^2 (\mathbf{k}_1 +\mathbf{k}_2+\mathbf{k}_3)^2}{\mathbf{q}_4^2\mathbf{k}_1^2}
\ln \frac{\mathbf{q}_2^2 (\mathbf{k}_1 +\mathbf{k}_2+\mathbf{k}_3)^2}{\mathbf{q}_1^2(\mathbf{k}_2 +\mathbf{k}_3)^2}
\right),  \nonumber
\eeqn

The remainder function $ R^{(2)}_{2 \to 5}$ has an interesting property of being expressed through the corresponding remainder function for the $2 \to 4 $ amplitude found in Ref.~\cite{BLS2}
\beqn
R^{(2)}_{2 \to 4} =-\frac{i \pi }{2} \ln \left(\frac{s_2}{s_0}\right) f_{6} (w, w^*),
\eeqn
where
\beqn\label{f6}
f_{6} (w, w^*)=\ln |1+w|^2 \ln \left| 1+\frac{1}{w}\right|^2.
\eeqn

We can compactly write the expression in eq.~(\ref{R25L2}) as a sum
\beqn\label{R25compact}
R^{(2)}_{2 \to 5}= \frac{i \pi }{ 2} \ln \sqrt{u_{25} \; u_{36} \; u_{51} \; u_{62}}
\left(f_{6} (w_a, w_a^*)+f_{6} (w_b, w_b^*)\right),
\eeqn
where
\beqn
w_a=\frac{w_1}{1+\frac{1}{w_2}}, \;\;\; w_b=\frac{1}{w_2} \frac{1}{1+w_1}.
\eeqn

In the next section we generalize the present discussion to  the  $2 \to 2 +(n-4 )$ scattering amplitudes with an arbitrary  number  of produced gluons $n-4$.

\section{Some generalizations for more legs}\label{twoloopsmorelegs}

\begin{figure}[h]
  \begin{center}
    %\showthe\columnwidth % Use this to determine the width of the figure.
    \includegraphics[width=0.35\columnwidth]{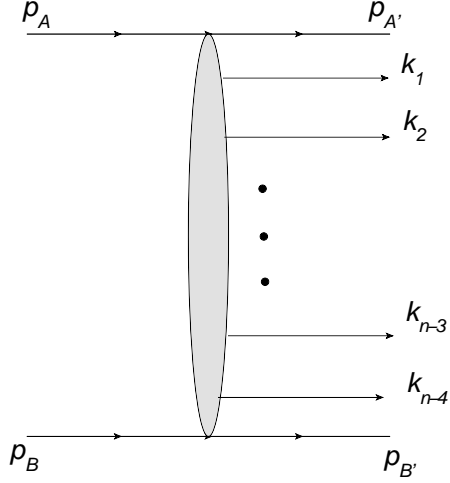}
    \caption{ The $2 \to 2+ (n-4)$ scattering amplitude with $n-4$ produced gluons. }\label{fig:multi2NN}
  \end{center}
\end{figure}

\begin{figure}[h]
  \begin{center}
    %\showthe\columnwidth % Use this to determine the width of the figure.
    \includegraphics[width=0.15\columnwidth]{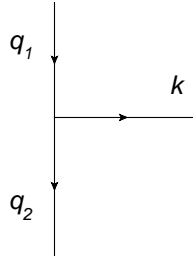}
    \caption{ The effective reggeon-particle-reggeon emission vertex. The $t$-channel gluons are not  yet dressed at the Born approximation and thus are denoted by simple~(not curvy) lines. }
    \label{fig:vertex}
  \end{center}
\end{figure}

In this section we  apply the results of the previous section  to the $2 \to 2+(n-4)$    scattering amplitudes with  $n-4$ produced gluons  illustrated in Fig.~\ref{fig:multi2NN}. We argue that at two loops the corresponding remainder  function in the Mandelstam regions, where we flip at least two adjacent particles can be written as a linear combination of the $2 \to 4$ remainder function.
This happens due to a very special structure of the effective emission  for a definite helicity vertex of the produced particles
\beqn\label{vertex}
\sqrt{2} \frac{q_1 q_2^*}{k}
\eeqn
shown in Fig.~\ref{fig:vertex}.

If we consider an effective diagram with emission of two gluons with the  same helicity, when the transverse propagator between the two emissions is canceled as follows
\beqn
\sqrt{2}  \; \frac{q_1 q_2^*}{k^*_1}\frac{1}{|q_2|^2}
\sqrt{2}\; \frac{q_2 q_3^*}{k^*_2}=2   \frac{q_1 q_3^*}{k^*_1 k_2^*}.
\eeqn

Using this property we can  find the impact factor $\tilde{\Phi}_{34}$ with two emissions  for the discontinuity in $s$, similar to $\tilde{\Phi}_2$   of eq.~(\ref{Phi2tilde}).
Let us recall how one  calculates impact factor $\tilde{\Phi}_2$ depicted in Fig.~\ref{fig:F2tilde}. Plugging in the effective emission vertex and the transverse propagator we write
\beqn
\sqrt{2} \frac{(k''-k_2)(k''-k_2-k_3)^*}{k_3^*} \frac{1}{|k''-k_2-k_3|^*}=
\sqrt{2}\frac{(k''-k_2)}{k_3^* (k''-k_2-k_3)}
\eeqn
and then divide this  by Born expression
\beqn
\sqrt{2} \frac{(q_2-k_2) (q_2-k_2-k_3)^*}{k_3^*} \frac{1}{|q_2-k_2-k_3|^2}
=\sqrt{2} \frac{(q_2-k_2) }{k_3^* (q_2-k_2-k_3)}.
\eeqn
As a result we have
\beqn
 \tilde{ \Phi_2}=\frac{(q_2-k_2-k_3) (k''-k_2)}{(q_2-k_2) (k''-k_2-k_3)}=
 \frac{q_4 (k''-k_2)}{q_3 (k''-k_2-k_3)}.
\eeqn

\begin{figure}[h]
  \begin{center}
    %\showthe\columnwidth % Use this to determine the width of the figure.
    \includegraphics[width=0.4\columnwidth]{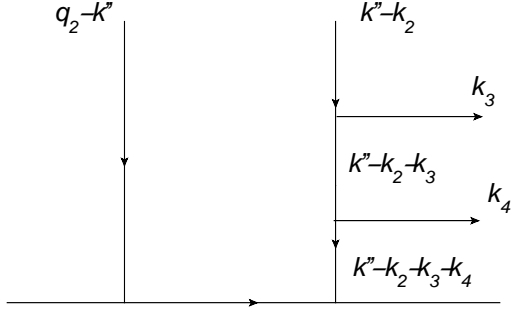}
    \caption{ Impact factor for the discontinuity in $s$ with two produced gluons.  }
    \label{fig:F23tilde}
  \end{center}
\end{figure}

\begin{figure}[h]
  \begin{center}
    %\showthe\columnwidth % Use this to determine the width of the figure.
    \includegraphics[width=0.65\columnwidth]{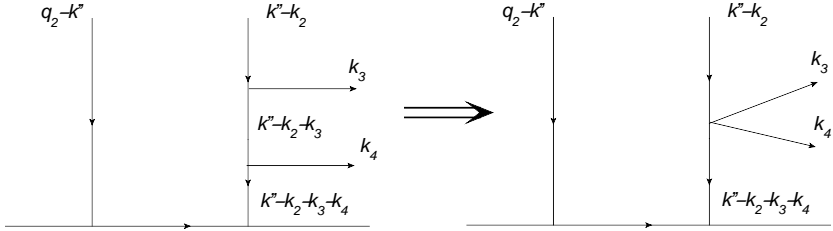}
    \caption{ The impact factor $\tilde{\Phi}_{34}$ for two produced gluons of the same helicity with momenta $k_3$ and $k_4$ can be written as an impact factor with one produced gluon with momentum $k_3+k_4$.  }
    \label{fig:F23tildeRed}
  \end{center}
\end{figure}

\begin{figure}[h]
  \begin{center}
    %\showthe\columnwidth % Use this to determine the width of the figure.
    \includegraphics[width=0.65\columnwidth]{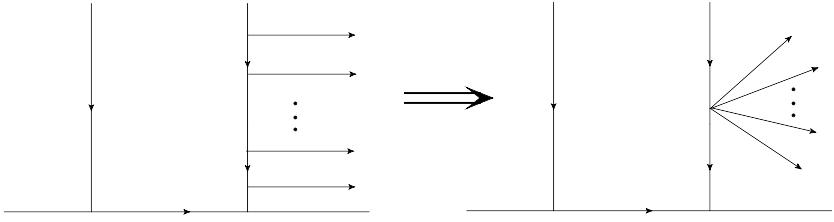}
    \caption{ The impact factor $\tilde{\Phi}_{i...j}$ for $i+...+j$ produced gluons of the same helicity with momenta $k_i$, $k_{i+1}$, ...  ,$k_j$ can be written as an impact factor with one produced gluon with momentum $k_i+...+k_j$.}
    \label{fig:Fn}
  \end{center}
\end{figure}

\begin{figure}[h]
  \begin{center}
    %\showthe\columnwidth % Use this to determine the width of the figure.
    \includegraphics[width=0.65\columnwidth]{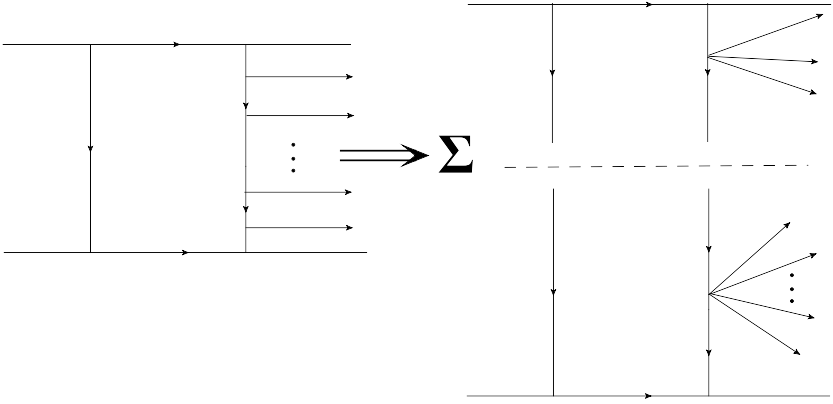}
    \caption{ The $2 \to 2+(n-4)$ scattering amplitude can be factorized into two pieces, which are then summed over. The dashed line denotes the factorization point. }
    \label{fig:sum}
  \end{center}
\end{figure}

Next we introduce one more produced gluon $k_4$ as illustrated in Fig.~\ref{fig:F23tilde}
and calculate the corresponding expression by plugging in the effective vertices and the transverse gluon propagators
\beqn
&&\sqrt{2}\frac{(k''-k_2)(k''-k_2-k_3)^*}{k_3^*} \frac{1}{|k''-k_2-k_3|^2}
\sqrt{2} \frac{(k''-k_2-k_3) (k''-k_2-k_3-k_4)}{k^*_4} \frac{1}{|k''-k_2-k_3-k_4|^2} \;\;\;\; \nonumber
\\
&&
 \hspace{2cm}=2 \frac{k''-k_2}{k''-k_2-k_3-k_4} \frac{1}{k^*_3 k^*_4}.
\eeqn
Dividing this by the corresponding Born expression
\beqn
&&
\sqrt{2} \frac{(q_2-k_2) (q_2-k_2-k_3)^*}{k_3^*} \frac{1}{|q_2-k_2-k_3|^2} \sqrt{2} \frac{(q_2-k_2-k_3)(q_2-k_2-k_3-k_4)^*}{k_4^*} \frac{1}{|q_2-k_2-k_3-k_4|^2} \nonumber
\\
&&
\hspace{2cm}= 2\frac{q_2-k_2}{q_2-k_2-k_3-k_4} \frac{1}{k_3^* k_4^*}
\eeqn
we readily get
\beqn\label{Phi34}
\tilde{\Phi}_{34}=
\frac{(q_2-k_2-k_3-k_4) (k''-k_2)}{(q_2-k_2) (k''-k_2-k_3-k_4)}=
 \frac{q_5 (k''-k_2)}{q_3 (k''-k_2-k_3-k_4)}.
\eeqn
An important observation is in order. The impact factor   for an emission of two gluons with the same helicity can be obtained from a  corresponding impact factor with one gluon emission  by shifting the transverse momentum of the produced gluon. In other words impact factor $\tilde{\Phi}_{34}$ is the same as  $\tilde{\Phi}_{2}$ with $k_3 \to k_3+k_4$. Graphically this statement means that any two adjacent gluon emissions of the same helicity can be represented as a stretched diagram in the transverse space, where the two gluons are emitted at the same point as shown in Fig.~\ref{fig:F23tildeRed}.
In a more general case of many gluon emissions of the same helicity all of them effectively emitted from the same point due to the cancelation of the propagators between them as shown in Fig.~\ref{fig:Fn}.
The corresponding impact factor $\tilde{\Phi}_{m}$ for $m$ emitted gluons can be obtained from  $\tilde{\Phi}_{2}$ in eq.~(\ref{Phi2tilde}) by  shifting   the transverse momentum $k_3 \to k_4 +...+k_m$.
This fact allows us  to factorize any $2 \to 2 +(n-4)$ amplitude  into two pieces with redefined impact factors. These two new redefined impact factors are then convolved with the BFKL propagators in a way it was done  for the  $2 \to 4$ amplitude and thus give  the $2 \to 4$  answer with redefined momenta. The number of factorization points determines the number of $2 \to 4$-like terms in the final answer. For Mandelstam region of the $2 \to  2+ (n-4)$ scattering amplitude, where all $n-4$ produced particles are flipped we have $n-5$ factorization points denoted by a dashed line in Fig.~\ref{fig:sum}. In the case of the $2 \to 5 $ amplitude we have two factorization points, which explains why the two-loop remainder function in eq.~(\ref{R25compact}) can be written as a linear combination of two $2 \to 4$ remainder function.

It is worth emphasizing that here we considered impact factors $\tilde{\Phi}_i$ for the discontinuity in $s$. We showed~(see eq.~(\ref{PhiPhitilde}) and text wherein) that in the case of the $2 \to 5$ amplitude they differ from those for discontinuity in $s_{123}$ only by  an additive number
\beqn
\Phi_i+ \tilde{\Phi}_i=1.
\eeqn
The difference between them is important only at one loop level and changes the regularization prescription. Starting at two loops it vanishes after the integration over the loop momenta because of the BFKL propagators.
The best way to see this is to look at the integral representation eq.~(\ref{disc}), which is divergent at one loop at $\nu=0$~($\mu=0$) for $n=0$~($m=0$), while starting at two loops this divergency is absent due  to the fact that the BFKL eigenfunctions $E_{\nu,n}$ vanish at those points.
  Any emission of one or more produced gluons does not change this property  and the self consistency requires the remainder function for the $2 \to 2+(n-4)$  MHV amplitude in this Mandelstam region calculated from the  discontinuities in $s_{1...(n-4)}$ and $s$ to be the same.

This simple iterative structure breaks down if we go beyond two loops. However, we expect   a  similar, though   more  complicated iterations at higher loops, which will be published by us elsewhere.

At two loops we can formulate a  simple mnemonic rule based on the iterative property of the impact factors outlined above. For any given $2 \to 2 +(n-4)$ scattering MHV amplitude in the Mandelstam region, where  all $n-4$ produced particles are flipped,  the amplitude can be factorized into two pieces in $n-5$ ways as shown in Fig.~\ref{fig:sum}. The dashed line denotes the factorization point. Let us consider the upper piece in Fig.~\ref{fig:sum}  in more details. We insert the completeness condition eq.~(\ref{compBFKL}) for the BFKL eigenfunction in the factorization point and integrate over the transverse momenta in a similar way we did  in eq.~(\ref{chi2}). The impact factor $\tilde{\Phi}_{123}$  in $\nu, n$ space for the emission of three produced gluons with momenta $k_1$, $k_2$ and $k_3$  depends only on the following ratio of the complex transverse momenta
\beqn
\frac{q_1}{k_1+k_2+k_3}.
\eeqn
 In a similar way for the lower piece in Fig.~\ref{fig:sum}  for $n-7$ produced particles  with momenta $k_4$, ..., $k_{n-5}$ and  $k_{n-4}$  one has (note inverse momenta ordering)
 \beqn
\frac{k_4+k_5+...+k_{n-4}}{q_{n-3}}.
\eeqn
The product of two impact factors depends only on the complex cross ratio in the space of the transverse momenta
\beqn
w_{3}=\frac{q_1(k_4+k_5+...+k_{n-4})}{q_{n-3} (k_1+k_2+k_3)},
\eeqn
where the subindex of $w_{3}$ stands for the factorization point between gluons with momenta $k_3$ and $k_4$.
In general, it is given by
\beqn
w_i=\frac{q_1 (k_{i+1}+...+k_{n-4})}{q_{n-3} (k_1+...+k_i)}.
\eeqn
Then the two loop remainder function for the $2 \to 2 +(n-4)$ scattering amplitude in the Mandelstam region, where we flip all of the  $n-4$ produced particles reads
\beqn\label{R2n4}
R^{(2)}_{2 \to 2 +(n-4) }= -\frac{i\pi}{2} \ln \left( \frac{s_{1...(n-4)}}{s_0} \right) \sum_{i=1}^{n-5} f_{6}(w_i, w_i^*).
\eeqn
The function of the transverse momenta  $f_{6} (w, w^*)$ appears in the remainder function for $2 \to 4$ scattering amplitude and  was defined in eq.~(\ref{f6}). Using the transverse momenta  conservation
\beqn
q_1=q_{n-4}+k_1+...+k_{n-4}
\eeqn
we write (see Appendix~\ref{app:wi})
\beqn
f_6 (w_i,w^*_i)=
\ln|1+w_i|^2 \ln \left|1+\frac{1}{w_i}\right|^2
=
\ln \frac{\mathbf{q}^2_{i+1}(\mathbf{k}_1+..+\mathbf{k}_{n-4})^2}{\mathbf{q}^2_{n-3} (\mathbf{k}_1+...+\mathbf{k}_i)^2}
\ln \frac{\mathbf{q}^2_{i+1} (\mathbf{k}_1+..+\mathbf{k}_{n-4})^2}{\mathbf{q}^2_{1} (\mathbf{k}_{i+1}+...+\mathbf{k}_{n-4})^2}. \;\;\;\;
\eeqn

In the Mandelstam region under consideration, where we flip all produced particles, only one cross ratio processes the phase $-i2 \pi$, namely the one that has both $s$ and $s_{1...n-4}$ in the numerator. It is easy to identify it because it is goes to unity with the fastest rate in the multi-Regge kinematics
\beqn
1-U \simeq \delta^{n-5},
\eeqn
where
\beqn\label{delta}
\delta=\frac{\alpha_{i+1}}{\alpha_i}
\eeqn
is defined in terms of the Sudakov variables $\alpha_i$ in eq.~(\ref{Sudakov}).

In the next section we discuss  other Mandelstam regions of the $2 \to 2+(n-4)$  scattering amplitude at two loops.

 \subsection{Other Mandelstam regions at two loops in LLA }

 We can also consider   Mandelstam regions of the $2 \to 2+ (n-4)$ scattering amplitude,
 where we flip less than $n-4$ adjacent produced gluons having the same helicity. It is worth emphasizing that this discussion is not limited to MHV amplitudes and the only condition is that the produced gluons we flip  must have the same helicity.

 Any gluon emission of any helicity that does not participate in building the Mandelstam cut is factorized out and canceled when we divide the impact factor for the corresponding  Born expression. To illustrate this statement let us consider $2 \to 6 $ scattering amplitude in Fig.~\ref{fig:multi2NN} in the Mandelstam region where we flip first three produced particles with momenta $k_1$, $k_2$ and $k_3$. In this region we find the discontinuity in $s_{123}$ and, using the notation of the previous section, we can have only two factorization points between $k_1$ and $k_2$, and between $k_2$ and $k_3$. The factorization point between $k_3$ and $k_4$ is excluded because it does not correspond to the discontinuity in $s_{123}$.

\begin{figure}[h]
  \begin{center}
    %\showthe\columnwidth % Use this to determine the width of the figure.
    \includegraphics[width=0.35\columnwidth]{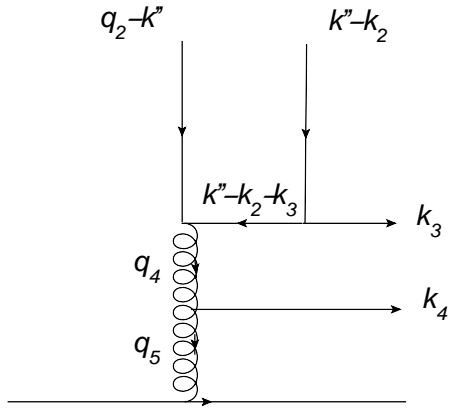}
    \caption{The impact factor with emission of two gluons with momenta $k_3$ and $k_4$. The helicity of the produced gluons is not necessarily the same.  }
    \label{fig:lowPhi34}
  \end{center}
\end{figure}

\begin{figure}[h]
  \begin{center}
    %\showthe\columnwidth % Use this to determine the width of the figure.
    \includegraphics[width=0.23\columnwidth]{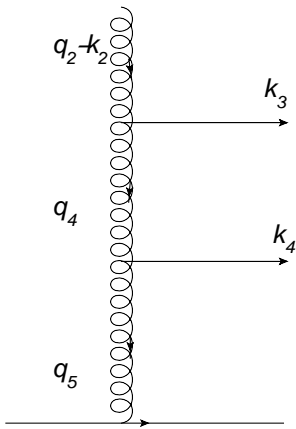}
    \caption{The Born expression for the two gluon production. The $t$-channel gluons are not yet reggeized in the Born approximation and the curvy lines are introduced merely to distinguish between reggeons and particles.  }
    \label{fig:Bron34}
  \end{center}
\end{figure}

  Let us look closely  at the lower impact factor for the factorization point between gluons with momenta  $k_2$ and $k_3$ shown in Fig.~\ref{fig:lowPhi34}.
Due to the Regge factorization   any dependence on $k_4$ and $q_5$ in the impact factor in Fig.~\ref{fig:lowPhi34} cancels if we divide it by the corresponding expression in the Born approximation depicted in Fig.~\ref{fig:Bron34}. As a result  we get back the same expression we had for $\Phi_2$ in Fig.~\ref{fig:F2}.

This means that to the leading order the impact factor that stands for Fig.~\ref{fig:lowPhi34} scaled by its Born expression  is  exactly the same as $\Phi_2$ that corresponds to Fig.~\ref{fig:F2}. This statement is independent on the helicity of the produced gluon with momentum $k_4$ and valid for any number of gluons of any helicity produced below  the one with momentum $k_3$. The same is true if we add gluon emissions above the first gluon that spans the energy for which we calculate the discontinuity of the amplitude.

Thus  we can  formulate a leading logarithmic prediction for the remainder function for an arbitrary $ \text{MHV}$  $2 \to 2+(n-4)$ amplitude at two loops in the Mandelstam region, where we flip any $k-m$   adjacent produced particle
 \beqn\label{R2nmk}
R^{(2)}_{2 \to 2 +(n-4), k,m }=-\frac{ i\pi}{2} \ln \left( \frac{s_{k...m}}{s_0} \right)
 \sum_{i=k}^{m-1} f_{6}(w_i, w_i^*), \;\;\; k \geq 1, \;\;  m \leq n-4.
\eeqn

The indices $k$  and $m$ label the first and the last particles that span the energy $s_{k..m}$ for which we calculate the discontinuity. The only condition here  is that all of the flipped particles should  have the same helicity, while the helicity configuration of all other produced particles is arbitrary and does not effect  the remainder function in eq.~(\ref{R2nmk}). The phases of the cross ratios  for these Mandelstam regions are calculated case by case.

All results of this section are valid only in the leading logarithmic approximation. The next-to-leading corrections to the effective gluon emission vertex break the recursive properties of the impact factors.

\section{Conclusions}

We calculated the remainder function  for the $2 \to 5$ scattering MHV amplitude with leading logarithmic accuracy in the Mandelstam region where $s, s_2, s_3, s_{123}> 0$ that corresponds to  all three produced particles being  flipped as shown in Fig.~\ref{fig:25flipped3}. The result is given  in the integral representation of eq.~(\ref{R25L}) to any  loop order  and  calculated analytically at two loops~(see eq.~(\ref{R25compact})). We found that the two loop leading log remainder function in this Mandelstam region can be  compactly written as a sum of two remainder functions for the $2 \to 4$ scattering amplitude calculated in Ref.~\cite{BLS2}.  This  iterative structure happens due to the fact that the impact factor for two or more produced gluons  is expressed though the impact factor for one gluon  emission  with  a shifted transverse momentum. This property of the impact factors makes it possible to derive   the remainder function for the $2 \to 2 +(n-4)$ MHV amplitude in the Mandelstam region,  where all $n-4$ produced gluons are flipped and have the same helicity~(see eq.~(\ref{R2n4})).   In this region we do not have contributions of the Bartels-Kwiecinski-Praszalowicz (BKP) states, which appear in other regions for $n \geq 8$. In that sense the computations are the similar to the $n=6$ case. This explains why the $2 \to 2+(n-4)$ result can be written as a sum of $n-5$ LLA remainder functions for $n=6$ MHV amplitude.

 Furthermore we  consider other Mandelstam regions  of the $2 \to 2 +(n-4)$ amplitude, where we flip any number of adjacent produced gluons having the same helicity. The corresponding remainder functions at two loops are  given by eq.~(\ref{R2nmk}). The last result is also   valid  for $\text{N}^k \text{MHV}$ amplitudes,  where the  flipped particles have the same helicity, while the helicity of all other produced particles is arbitrary and does not effect the contribution of the Mandelstam cuts to the leading logarithmic accuracy.

\section{ Acknowledgments}

We thank   S.~Caron-Huot, B.~Dixon, V. S. Fadin, J.~Henn,  G.P. Korchemsky, E. M. Levin, A. Sabio Vera,
V. Schomerus, A. Sever, M. Spradlin, C.-I Tan, C. Vergu, P. Vieira and A. Volovich for
helpful discussions.  The work of A.~P. is supported in part by
the US National Science Foundation under grant PHY-064310.
The work of A.~K. is partially supported by the Marie Curie Grant PIRG-GA-2009-256313.

\newpage

\appendix

\setcounter{equation}{0}

\renewcommand{\theequation}{A.\arabic{equation}}

\section{Central emission block}\label{sec:block}

\begin{figure}[h]
  \begin{center}
    %\showthe\columnwidth % Use this to determine the width of the figure.
    \includegraphics[width=0.4\columnwidth]{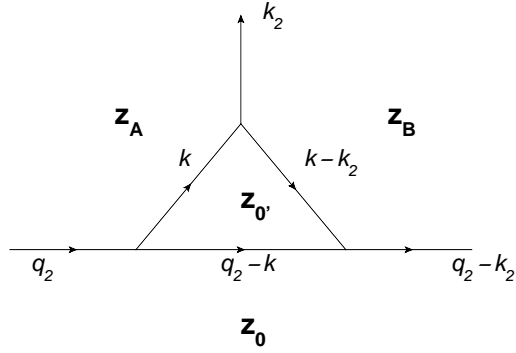}
    \caption{ Dual coordinates for the integral representation  of the central emission block. }\label{fig:triangledual}
  \end{center}
\end{figure}

In this section we calculate the central emission block shown in Fig.~\ref{fig:centerBKPL}. The dashed lines denote the insertion of the completeness condition for the BFKL eigenfunctions eq.~(\ref{compBFKL}) and the emission of the gluon with definite helicity  is given by the effective emission vertex  in eq.~(\ref{vertex}). Thus we can write for the central emission block in Fig.~\ref{fig:centerBKPL} the following integral
\beqn
\label{emission1}
J=\int \frac{d^2k}{\pi} \left(\frac{k^*}{q_2^*-k^*}\right)^{-i\nu +n/2}\,\left(\frac{k}{q_2-k}\right)^{-i\nu -n/2}
\frac{k (k^*-k_2^*)}{|k|^2|k-k_2|^2|q_2-k|^2 k^*_2}\left(\frac{k^*-k_2^*}{q_2^*-k^*}\right)^{i\mu -m/2}
\left(\frac{k-k_2}{q_2-k}\right)^{i\mu +m/2}.
\eeqn
It is useful to introduce the dual coordinates in the transverse momentum space to exploit the conformal properties of $J$. The dual coordinates are depicted in Fig.~\ref{fig:triangledual} and defined by
\beqn
&&  q_2-k_2=z_{B}-z_{0}, \;\; q_2=z_A -z_0, \;\; k_2=z_A-z_B, \;\; k=z_A-z_{0'},
\\
&&k-k_2= z_B-z_{0'} \;\; q_2-k=z_{0'}-z_{0}. \nonumber
\eeqn

The behavior of  the integral $J$ under inversion allows us to factor out the dependence on the external momenta
\beqn
&&
\int \frac{d^2 z_{0'}}{\pi} \left(  z^*_{A0'} \right)^{-i\nu +\frac{n}{2}-1}    \left(  z_{A0'} \right)^{-i\nu -\frac{n}{2}}
\left(  z^*_{B0'} \right)^{ i\mu -\frac{m}{2}} \left(  z_{B0'} \right)^{ i\mu +\frac{m}{2}-1}
\left(  z^*_{0'0} \right)^{i\nu -\frac{n}{2}-i\mu +\frac{m}{2}-1}
\left(  z_{0'0} \right)^{i\nu +\frac{n}{2}-i\mu -\frac{m}{2}} \hspace{1cm}\;\;\;  \\
&&
\xrightarrow[ ]{inversion}
(z_{0A})^{-i\mu -\frac{m}{2}}(z^*_{0A})^{-i\mu +\frac{m}{2}-1}
(z_{0B})^{i\nu +\frac{n}{2}-1}(z^*_{0B})^{i\nu -\frac{n}{2}}
(z_{AB})^{-i(\nu-\mu) -\frac{n-m}{2}}(z^*_{AB})^{-i(\nu-\mu) +\frac{n-m}{2}} \nonumber
\\
&&
\int \frac{d^2 z_{0'}}{\pi} \left(  z^*_{A0'} \right)^{-i\nu +\frac{n}{2}-1}    \left(  z_{A0'} \right)^{-i\nu -\frac{n}{2}}
\left(  z^*_{B0'} \right)^{ i\mu -\frac{m}{2}} \left(  z_{B0'} \right)^{ i\mu +\frac{m}{2}-1}
\left(  z^*_{0'0} \right)^{i\nu -\frac{n}{2}-i\mu +\frac{m}{2}-1}
\left(  z_{0'0} \right)^{i\nu +\frac{n}{2}-i\mu -\frac{m}{2}}.  \nonumber
\eeqn
Thus we can write
\beqn
 J=\frac{1}{|q_2|^2}
\frac{q_2 (q_2^*-k_2^*)}{k_2^*} \frac{1}{|q_2-k_2|^2}     \;\;
\left(\frac{k_2^*}{q_2^*-k_2^*}\right)^{-i\nu+\frac{n}{2}}\left(\frac{k_2}{q_2-k_2}\right)^{-i\nu-\frac{n}{2}}
\left(\frac{k_2}{q_2}\right)^{i\mu+\frac{m}{2}}\left(\frac{k_2^*}{q_2^*}\right)^{i\mu-\frac{m}{2}} \;\; B \;\;\;\;
\eeqn
The factor $B$ is a c-number, which is obtained by taking  $q_2 \to \infty$ and  $k_2=1 $ in  $J$
\beqn\label{CC}
&&B=\int \frac{d^2k}{\pi} (k^*)^{-i\nu +n/2-1}\,k^{-i\nu -n/2}
(k^*-1)^{i\mu -m/2}(k-1)^{i\mu +m/2-1}.
\eeqn
To calculate $B$ we introduce the  Sudakov variables
\beqn
k=k_1+i k_2=x, \;\;
k^*=k_1-i k_2=y
\eeqn
and write
\beqn
&&B=\int \frac{d x  d y  }{2\pi i}  \frac{x^{-n+1}}{ (x y -i\epsilon)^{i \nu -\frac{n}{2}+1}}
\frac{(x-1)^{m-1}}{((1-x)(1-y)-i\epsilon)^{-i\mu +\frac{m}{2}}}  \\
&&
=\int_0^1  d x  x^{-i\nu -\frac{n}{2}} (1-x)^{i\mu +\frac{m}{2}-1} (-1)^{m-1}
\frac{e^{-i\pi (-i \nu +\frac{n}{2}-1)} -e^{ i\pi (-i \nu +\frac{n}{2}-1)}}{2 \pi i } \nonumber
\\
&&
=
(-1)^{m-1}\frac{\Gamma (1-i\nu -n/2) }{\Gamma (1+i\nu -n/2)}\,
\frac{\Gamma (+i\mu +m/2) }{\Gamma (-i\mu +m/2)}\,\frac{\Gamma (i(\nu -\mu ) +(m-n)/2) }{\Gamma (1-i(\nu -\mu ) +(m-n)/2)}\nonumber
\eeqn

Then the central emission block $C$ is obtained by dividing $J$ by the corresponding Born expression
\beqn
\frac{1}{|q_2|^2}
\frac{q_2 (q_2^*-k_2^*)}{k_2^*} \frac{1}{|q_2-k_2|^2}
\eeqn
and reads
\beqn\label{Capp}
 C= %\frac{1}{|q_2|^2} \frac{q_2 (q_2^*-k_2^*)}{k_2^*} \frac{1}{|q_2-k_2|^2}     \;\;
-\frac{1}{2}\left(\frac{q_3^*}{k_2^*}\right)^{i\nu-\frac{n}{2}}\left(\frac{q_3}{k_2}\right)^{i\nu+\frac{n}{2}}
\left(\frac{q_2}{k_2}\right)^{-i\mu-\frac{m}{2}}\left(\frac{q_2^*}{k_2^*}\right)^{-i\mu+\frac{m}{2}} \;\; B.
\eeqn
 We normalize  $C$ in a way consistent with limits in eq.~(\ref{limits}).

\setcounter{equation}{0}

\renewcommand{\theequation}{B.\arabic{equation}}

\section{One loop from BDS in the Mandelstam region}\label{1loop}

In this section we consider the BDS formula at one loop for fixing the normalization of the remainder function $R_{2 \to 5}$. We perform the analytic continuation of the one loop expression to the Mandelstam region shown in Fig.~\ref{fig:25flipped3}, where we flip all produced particles with momenta $k_i$. In this Mandelstam region we have $s,s_2, s_3, s_{123}>0$ and  only one cross ratio that has a phase is
\beqn
u_{47}=|u_{47}| \;e^{-i2\pi}.
\eeqn

Firstly we decompose the BDS formula into two pieces - one corresponding  to Regge poles and the other coming from the beginning of the Mandelstam cut at one loop.
Generally, the piece that stands for Regge poles  for $2 \to 2+(n-4) $ amplitude  can be written as follows
\beqn\label{Rpoles}
&&
\Gamma_{PRP}(t_1)
 \left(\frac{-s_1}{\mu^2}\right)^{\omega(t_1)}\Gamma_{RPR} (t_1,t_2, \ln (-\kappa_1))
...
\left(\frac{-s_{i}}{\mu^2}\right)^{\omega(t_{i})}\Gamma_{RPR} (t_{i},t_{i+1}, \ln (-\kappa_{i}))
... \nonumber
\\
&&
 \left(\frac{-s_{n-4}}{\mu^2}\right)^{\omega(t_{n-4})}\Gamma_{RPR} (t_{n-4},t_{n-3}, \ln (-\kappa_{n-4}))
\left(\frac{-s_{n-3}}{\mu^2}\right)^{\omega(t_{n-3})}
\Gamma_{PRP}(t_{n-3}),
\eeqn
where the particle-reggeon-particle $\Gamma_{PRP}$ and  reggeon-particle-reggeon $\Gamma_{RPR}$ vertices were found  in Ref.~\cite{BLS1} from the BDS amplitude with $n=4$ and $n=5$ external gluons. To the first order in the coupling constant they are given by
 \beqn
\ln \Gamma_{PRP}(t)=-\frac{1}{\epsilon^2} +\frac{1}{2\epsilon}\ln \frac{-t}{\mu^2}+2 \zeta_2
\eeqn
and
\beqn
\ln \Gamma_{RPR}(t_1, t_2, \ln -\kappa)=-\frac{1}{2 \epsilon^2} -\frac{1}{4} \ln^2 \frac{-\kappa}{\mu^2} +\frac{1}{2} \ln \frac{-\kappa}{\mu^2} \left( \ln \frac{(-t_1)(-t_2)}{\mu^4} -\frac{1}{\epsilon}\right) -\frac{1}{4} \ln^2 \frac{-t_1}{-t_2}-\frac{1}{4}\zeta_2. \;\;\;\;\;\;
\eeqn
The gluon Regge trajectory is
\beqn
\omega(-\mathbf{q}^2)=a \left(\frac{1}{\epsilon} +\ln \frac{\mathbf{q}^2}{\mu^2}\right).
\eeqn
The parameters $\kappa_i$ depend only on transverse momenta in the multi-Regge kinematics  and for the $ 2 \to 5 $ amplitude  are given by

\beqn
-\kappa_1=\frac{(-s_1)(- s_2)}{(-s_{A'12})},
 \;\;
-\kappa_2=\frac{(-s_2)(- s_3)}{(-s_{123})},
\;\;
-\kappa_3=\frac{(-s_3)(- s_4)}{(-s_{23B'})}.
\eeqn

Then  we subtract from the   $n=7$\;\;\; BDS amplitude   the Regge pole contribution eq.~(\ref{Rpoles}) at one loop
 and get
\beqn\label{Fcuts}
&&F=\frac{1}{2 \epsilon} \ln u_{47} u_{73} u_{14}+ \frac{1}{2} \ln u_{73}  \ln \frac{(-s_2) \mu^2 }{(-t_1)(- t_3)}+\frac{1}{2} \ln u_{14} \ln \frac{(-s_3) \mu^2 }{(-t_2) (-t_4)}+\frac{1}{2}\ln u_{47} \ln  \frac{(-s_{123}) \mu^2 }{(-t_4)(-t_1)}\nonumber  \\
&& +\frac{\pi ^2}{3}-\frac{\ln^2 u_{14}}{4}-\frac{\ln^2 u_{47}}{4}-\frac{\ln^2 u_{73}}{4}-\frac{1}{2} \sum_{i=1}^{7} \text{Li}_2 (1-u_{i,i+3}).
\eeqn
The function $F$ carries the information about the Mandelstam cuts and is given in an arbitrary kinematics. This fact allows us to perform the analytic continuation of $F$ to the Mandelstam region where $u_{47}=|u_{47}| \;e^{-i2\pi}$ in the multi-Regge kinematics and obtain
\beqn\label{Fcont}
F \simeq -\frac{i \pi }{\epsilon }-i \pi \ln \frac{(\mathbf{k}_1+\mathbf{k}_2+\mathbf{k}_3)^2\mu^2}{\mathbf{q}^2_1\mathbf{q}^2_4}.
\eeqn
Note that $F \to 0$ for  $u_{14}, u_{73}, u_{47} \to 1$ in the multi-Regge kinematics.

\setcounter{equation}{0}

\renewcommand{\theequation}{C.\arabic{equation}}

\section{Integral at one and two loops}\label{sec:twoloops}
In this section we calculate   the integral in eq.~(\ref{R25L})
\beqn\label{Iell}
I^{(\ell)}=\sum_{n=-\infty}^{+\infty} \sum_{m=-\infty}^{+\infty} \int_{-\infty}^{+\infty} d \nu \int_{-\infty}^{+\infty} d \mu
 \;\;
\chi_1 \;\; C \;\; \chi_2  \;\; (E_{\nu,n}+E_{\mu,m})^{\ell-1}
\eeqn
for $\ell=1$ and $\ell=2$.

The integrand is given by
\beqn\label{chi1Cchi2}
 \chi_1 \;C\; \chi_2 =\frac{(-1)^{n+m}}{8}
\frac{\Gamma (-i\nu -\frac{n}{2}) }{\Gamma (1+i\nu -\frac{n}{2})}\,
\frac{\Gamma (i\mu +\frac{m}{2}) }{\Gamma (1-i\mu +\frac{m}{2})}\,\frac{\Gamma (i(\nu -\mu ) +\frac{m-n}{2}) }{\Gamma (1-i(\nu -\mu ) +\frac{m-n}{2})}
 w_1^{i\nu+\frac{n}{2}} (w^*_1)^{i\nu-\frac{n}{2}}
 w_2^{i\mu+\frac{m}{2}} (w^*_2)^{i\mu-\frac{m}{2}}, \nonumber
\eeqn
where $E_{\nu,n}$ is given by eq.~(\ref{Enun})
\beqn\label{Enunapp}
E_{\nu, n} =-\frac{1}{2} \frac{|n|}{\nu^2+\frac{n^2}{4}}
+\psi \left(1+i\nu +\frac{|n|}{2} \right)
+\psi \left(1-i\nu +\frac{|n|}{2} \right)
-2 \psi(1)
\eeqn
 and $w_i$ are expressed in terms of the complex transverse momenta
\beqn
w_1=\frac{k_1 q_3 }{ q_1 k_2}, \;\;\; w_2=\frac{k_2 q_4 }{ q_2 k_3}.
\eeqn
It is useful to introduce the phase   and the square of the absolute value   of $w_i$ as follows
\beqn
\alpha_i=\frac{w_i}{w_i^*}, \;\;\; \beta_i=|w|^2.
\eeqn
Firstly we calculate  $I^{\ell}$ at  one loop $\ell=1$
\beqn\label{I1}
&& I^{(1)}=\sum_{n=-\infty}^{+\infty} \sum_{m=-\infty}^{+\infty} \int_{-\infty}^{+\infty} d \nu \int_{-\infty}^{+\infty} d \mu
 \;\;
\frac{(-1)^{n+m}}{8}
\frac{\Gamma (-i\nu -\frac{n}{2}) }{\Gamma (1+i\nu -\frac{n}{2})}\,
\frac{\Gamma (i\mu +\frac{m}{2}) }{\Gamma (1-i\mu +\frac{m}{2})}\,
\\
&&  \hspace{2cm} \times \frac{\Gamma (i(\nu -\mu ) +\frac{m-n}{2}) }{\Gamma (1-i(\nu -\mu ) +\frac{m-n}{2})}
\alpha_1^{n/2} \beta_1^{i \nu}
\alpha_2^{m/2} \beta_2^{i \mu}.   \nonumber
\eeqn
 The calculation of the integral in eq.~(\ref{I1}) becomes much simpler if the consider only one region, where we have $\beta_1<1$ and $\beta_2>1$ which is consistent with  $w_1 \leftrightarrow 1/w_2$ symmetry. Moreover we focus on summing contributions of $\sqrt{\beta_i /\alpha_i}$ in that region and then exploit the $w_i  \leftrightarrow  w^*_i$ symmetry to restore the full answer.

The three gamma functions in the numerator of $I^{(1)}$ have  an infinite number of the simple poles. We show that only first two gamma functions contribute in the region under consideration and  that out of an infinite number of poles only one has non-vanishing residue. This is consistent with the $2 \to 4$ case, where  integrand that corresponds to one loop expression has only one simple pole.

The poles of the first gamma function in the numerator are located at
\beqn
-i\nu -\frac{n}{2}=-s, \;\;\; s=0,1,...
\eeqn
and the second gamma function has poles at
\beqn
i\mu +\frac{m}{2}=-t, \;\;\; t=0,1,...
\eeqn
Thus the contribution of the residues at these poles is given by
\beqn
\hspace{-1cm}\frac{\pi^2}{2} \sum_{n=-\infty}^{-1} \sum_{m=1}^{+\infty}
 \;\;
\frac{(-1)^{n+m} \;\; \Gamma (m-n+s+t) }{\Gamma (1-n+s) \Gamma (1-s-t) \Gamma (1+m+t)}\alpha_1^{n/2} \beta_1^{-n/2} \alpha_2^{m/2} \beta_2^{-m/2},   \;\;\;
\eeqn
where the limits of the summation are determined by the  convergency of the integrals over $\nu$ and $\mu$ at the  large circle in the complex planes for $\beta_1<1$ and $\beta_2 >1$.  There are also contributions at $n=0$ and $m=0$, which we consider separately.
We see that $\Gamma (1-s-t)$ makes this expression to vanish for all $s$ and $t$ except $s=t=0$ and therefore it becomes
\beqn\label{mnneq0}
&& \frac{\pi^2}{2}
 \sum_{n=-\infty}^{-1} \sum_{m=1}^{+\infty}
 \;\;
\frac{(-1)^{n+m} \;\; \Gamma (m-n) }{\Gamma (1-n) \Gamma (1+m)}\alpha_1^{n/2} \beta_1^{-n/2} \alpha_2^{m/2} \beta_2^{-m/2}=
\frac{\pi^2}{2} \ln \left(1+\sqrt{\frac{\beta_1}{\alpha_1}}\right)
\\
&&
+\frac{\pi^2}{2} \ln \left(1+\sqrt{\frac{\alpha_2}{\beta_2}}\right)
-\frac{\pi^2}{2}  \ln \left(1+\sqrt{\frac{\beta_1}{\alpha_1}}+\sqrt{\frac{\alpha_2}{\beta_2}}\right).
\nonumber
\eeqn
Next we find the corresponding terms for $n=0, m\neq 0$
\beqn
 \frac{\pi^2}{2}  \sum_{m=1}^{+\infty}
 \;\;
\frac{(-1)^{m} \;  \Gamma (m) }{ \Gamma (1+m)} \;  \alpha_2^{m/2} \beta_2^{-m/2}=-\frac{\pi^2}{2} \ln \left(1+\sqrt{\frac{\alpha_2}{\beta_2}}\right)
\eeqn
and  for $m=0, n\neq 0$
\beqn
 \frac{\pi^2}{2}  \sum_{n=-\infty}^{-1}
 \;\;
\frac{(-1)^{n} \;  \Gamma (-n) }{ \Gamma (1-n)} \;  \alpha_1^{n/2} \beta_1^{-n/2}=-\frac{\pi^2}{2} \ln \left(1+\sqrt{\frac{\beta_1}{\alpha_1}}\right).
\eeqn
Adding these two to eq.~(\ref{mnneq0}) we obtain
\beqn
-\frac{\pi^2}{2}  \ln \left(1+\sqrt{\frac{\beta_1}{\alpha_1}}+\sqrt{\frac{\alpha_2}{\beta_2}}\right)
\eeqn

Going back to the integrand in eq.~(\ref{chi1Cchi2})
we see that we have also poles from the third gamma function in the numerator, namely from $\Gamma (i(\nu -\mu ) +\frac{m-n}{2})$. These poles lead to a  system of inequalities for the arguments  of the gamma functions  in the denominator  of   the integrand in eq.~(\ref{chi1Cchi2})  that comes from the requirement of the arguments being greater than zero for non-vanishing integrand. Those inequalities have no solution in the region under consideration.
Going back to the complex variables $w_i$ we write
\beqn
 -\frac{\pi^2}{2}  \ln \left(1+\sqrt{\frac{\beta_1}{\alpha_1}}+\sqrt{\frac{\alpha_2}{\beta_2}}\right)
=
-\frac{\pi^2}{2}  \ln \left(1+w_1^*+\frac{1}{w_2^*}\right),
\eeqn
and  the symmetrization in $w_i \leftrightarrow w_i^*$ gives
\beqn\label{I1nmneq0}
I^{(1)}_{m=n\neq 0}=-\frac{\pi^2}{2}  \ln \left|1+w_1+\frac{1}{w_2}\right|^2
=
-\frac{\pi^2}{2}  \ln \frac{\mathbf{q}^2_2 \mathbf{q}^2_3 ( \mathbf{k}_1+\mathbf{k}_2+\mathbf{k}_3)^2}{\mathbf{q}^2_1 \mathbf{q}_4^2 \mathbf{k}_2^2}.
\eeqn
For $n=m=0$ the integral $I^{(1)}$ diverges at $\nu,\mu=0$ and need to be regularized. For $\beta_1 <1$ and $\beta_2 >1$    we choose a regularization that is compatible with the BDS amplitude as follows. By comparison with eq.~\ref{I1nmneq0} we get
\beqn
\text{Reg}_{s_{123}}\int^{+\infty}_{-\infty}  d \nu \int^{+\infty}_{-\infty}  d \mu
\frac{\beta_1^{i\nu} \;\;\;\beta_2^{i \mu}}{8}
\frac{\Gamma(i \mu) \Gamma(-i\nu ) \Gamma(i \nu-i\mu)}{\Gamma(1-i\mu) \Gamma(1+i\nu) \Gamma(1-i\nu+i \mu)}= \frac{\pi^2}{2} \left(-\frac{i\pi}{\epsilon}+\ln \frac{\mathbf{q}_2^2\mathbf{q}_3^2}{\mathbf{k}_2^2  \;\mu^2}\right),\;\;\;\;\;\;
\eeqn
which is in an agreement with the regularization for the $2 \to 4$ amplitude found in Ref.~\cite{BLS2}.

Next we calculate $I^{(\ell)}$ at two  loops
 \beqn\label{I2}
&& I^{(2)}=\sum_{n=-\infty}^{+\infty} \sum_{m=-\infty}^{+\infty} \int_{-\infty}^{+\infty} d \nu \int_{-\infty}^{+\infty} d \mu
 \;\;
\frac{(-1)^{n+m}}{8}
\frac{\Gamma (-i\nu -\frac{n}{2}) }{\Gamma (1+i\nu -\frac{n}{2})}\,
\frac{\Gamma (i\mu +\frac{m}{2}) }{\Gamma (1-i\mu +\frac{m}{2})}\,
\\
&&  \hspace{2cm} \times \frac{\Gamma (i(\nu -\mu ) +\frac{m-n}{2}) }{\Gamma (1-i(\nu -\mu ) +\frac{m-n}{2})}
\alpha_1^{n/2} \beta_1^{i \nu}
\alpha_2^{m/2} \beta_2^{i \mu} \left(E_{\nu,n} +E_{\mu,m}\right).   \nonumber
\eeqn

The two loop expression $I^{(2)}$ is finite at $n=m=0$ and $\nu,\mu=0$ due to the fact that  the BFKL eigenvalues $E_{\nu,n}$ and $E_{\mu,m}$ in eq.~(\ref{Enunapp})   vanish at those points. We also note that the symmetry $n \leftrightarrow -m$, $\nu \leftrightarrow -\mu$ of $I^{(2)}$ corresponds to  $w_1 \leftrightarrow 1/w_2 $. Thus it is enough to find only the contribution of $E_{\nu,n}$ and then that of $E_{\mu,m}$ in eq.~(\ref{I2}) is obtained by symmetrizing the result with respect to  $w_1 \leftrightarrow 1/w_2 $. Moreover taking into account that the result should be symmetric in   $w_i \leftrightarrow w^*_i$,  it is  enough to calculate only the piece that depends on $w_i^*=\sqrt{\beta_i/\alpha_i}$. As in the case of the one loop calculation we pick up the  region that is compatible with $w_1 \leftrightarrow 1/w_2 $ symmetry, namely $\beta_{1} <1, \;\;\;\beta_2 >1$. We calculate
\beqn\label{I2Enun}
&&
\sum_{n=-\infty}^{+\infty} \sum_{m=-\infty}^{+\infty} \int_{-\infty}^{+\infty} d \nu \int_{-\infty}^{+\infty} d \mu
 \;\;
\frac{(-1)^{n+m}}{8}
\frac{\Gamma (-i\nu -\frac{n}{2}) }{\Gamma (1+i\nu -\frac{n}{2})}\,
\frac{\Gamma (i\mu +\frac{m}{2}) }{\Gamma (1-i\mu +\frac{m}{2})}\,
\\
&&  \hspace{2cm} \times \frac{\Gamma (i(\nu -\mu ) +\frac{m-n}{2}) }{\Gamma (1-i(\nu -\mu ) +\frac{m-n}{2})}
\alpha_1^{n/2} \beta_1^{i \nu}
\alpha_2^{m/2} \beta_2^{i \mu}  E_{\nu,n}. \nonumber
\eeqn
The integration over  $\mu$ and  the summation  over $m$ is not difficult to do if we recall that at one loop all poles that have non-vanishing residue in the our region come from   $\Gamma (i\mu +\frac{m}{2}) $, while all residues of poles  of $\Gamma (i(\nu -\mu ) +\frac{m-n}{2})$ in $\mu$ are zero. The integrand of eq.~(\ref{I2Enun}) differs from that of the one loop integral in eq.~(\ref{I1}) only by $E_{\nu,n}$, which does not effect the integration over  $\mu$. As we saw in the one loop case, the poles of $\Gamma (i\mu +\frac{m}{2}) $ are located at $i\mu +\frac{m}{2}=-t, \;\;\;t=0,1,...$, but  due to the very special structure of the integrand only pole with $t=0$ contribute. Using the Cauchy theorem we integrate eq.~(\ref{I2Enun}) over $\mu$ and obtain
 \beqn\label{I2Enun3}
 && 2\pi \sum_{n=-\infty}^{+\infty} \sum_{m=0}^{+\infty} \int_{-\infty}^{+\infty} d \nu
 \;\;
\frac{(-1)^{n+m}}{8}
\frac{\Gamma (-i\nu -\frac{n}{2}) }{\Gamma (1+i\nu -\frac{n}{2})}\,
\frac{\Gamma (m-\frac{n}{2}+i\nu) }{\Gamma (1-i \nu  -\frac{ n}{2})}
\frac{\alpha_1^{n/2} \beta_1^{i \nu}
\alpha_2^{m/2} \beta_2^{-m/2} }{\Gamma (1+m)}\,
  E_{\nu,n} \hspace{1cm} \;\;\;\;
\\
&&
\hspace{2cm} =\frac{\pi}{4} \sum_{n=-\infty}^{+\infty} \int_{-\infty}^{+\infty} d \nu   \frac{(-1)^n \alpha_1^{\frac{n}{2}} \beta_1^{i\nu}}{\nu^2+\frac{n^2}{4}} \;E_{\nu,n}  \left(1+\sqrt{\frac{\alpha_2}{\beta_2}}\right)^{-i\nu +\frac{n}{2}}.  \nonumber
\eeqn
The expression in eq.~(\ref{I2Enun3}) is reduced to the two loop integral for the $n=6$ remainder function
calculated in Ref.~\cite{BLS2}( for $\beta_1 <1$ and $\beta_2 >1$)
\beqn
\frac{\pi}{4} \sum_{n=-\infty}^{+\infty} \int_{-\infty}^{+\infty} d \nu   \frac{(-1)^n \tilde{\alpha}^{\frac{n}{2}} \tilde{\beta}^{i\nu}}{\nu^2+\frac{n^2}{4}} \;E_{\nu,n}
=
\frac{\pi^2}{4} \ln^2 \left( 1+\sqrt{\frac{\tilde{\beta}}{\tilde{\alpha}}}\right)
-\frac{\pi^2}{4} \ln \left( 1+\sqrt{\frac{\tilde{\beta}}{\tilde{\alpha}}}\right) \ln \tilde{\beta}
\eeqn
with redefined variables
\beqn
\tilde{\alpha}=\alpha_1  \left(1+\sqrt{\frac{\alpha_2}{\beta_2}}\right),
\;\;\;
\tilde{\beta}= \frac{\beta_1}{ 1+\sqrt{\frac{\alpha_2}{\beta_2}} },
\eeqn
which correspond to
\beqn
\tilde{w}^* =\frac{w_1^*}{1+\frac{1}{w_2^*}}.
\eeqn
Symmetrizing the result in $w_i \leftrightarrow w_i^*$ we get
\beqn\label{I2Enun}
&&
\sum_{n=-\infty}^{+\infty} \sum_{m=-\infty}^{+\infty} \int_{-\infty}^{+\infty} d \nu \int_{-\infty}^{+\infty} d \mu
 \;\;
\frac{(-1)^{n+m}}{8}
\frac{\Gamma (-i\nu -\frac{n}{2}) }{\Gamma (1+i\nu -\frac{n}{2})}\,
\frac{\Gamma (i\mu +\frac{m}{2}) }{\Gamma (1-i\mu +\frac{m}{2})}\,
\\
&&  \hspace{1cm} \times \frac{\Gamma (i(\nu -\mu ) +\frac{m-n}{2}) }{\Gamma (1-i(\nu -\mu ) +\frac{m-n}{2})}
\alpha_1^{n/2} \beta_1^{i \nu}
\alpha_2^{m/2} \beta_2^{i \mu}  E_{\nu,n}
 \nonumber
=\frac{\pi^2}{4} \ln \left|1+\tilde{w} \right|^2\ln \left|1+\frac{1}{\tilde{w}} \right|^2
\eeqn
The contribution of $E_{\mu,m}$ is obtained by substitution $w_1 \leftrightarrow 1/w_2$ and we finally get  the two loop answer for the $n=7$ amplitude in terms of the function $f_6(w,w^*)$ that appears in $n=6$ amplitude
\beqn
I^{(2)}= \frac{\pi^2}{4} \left(f_6(w_a,w_a^*)+f_6(w_b,w_b^*)\right),
\eeqn
where
\beqn
f_6(w,w^*)=\ln \left|1+w \right|^2\ln \left|1+\frac{1}{w} \right|^2
\eeqn
and
\beqn
w_a=\frac{w_1}{1+\frac{1}{w_2}}, \;\;\; w_b=\frac{1}{w_2} \frac{1}{1+w_1}.
\eeqn

\setcounter{equation}{0}

\renewcommand{\theequation}{D.\arabic{equation}}

\section{Larger number of external gluons}\label{app:wi}

In the previous section we found  that the $n=7$ remainder function can be expressed in terms of the $n=6$ remainder function to the leading logarithmic accuracy in the Mandelstam region under consideration. The energy dependence  is known and the problem is reduced to calculating  the finite function of  transverse momenta. In section~\ref{twoloopsmorelegs} we  showed that the leading order impact factors have a recursive properties for emissions of the gluons with the same helicity. This happens due to the effective cancelation of the transverse propagators between the emitted gluons,  and  thus the impact factor with $m$ emitted gluons can be written as an impact factor for one gluon with shifted  transverse momentum $k_i+k_{i+1}+...+k_m$. This property allows us to factorize $2 \to 2+(n-4)$ amplitude with $n-4$ produced gluons in $n-5$ ways. In the Mandelstam region, where we flip all of the produced gluons we have
\beqn\label{wi}
w_i=\frac{q_1 (k_{i+1}+...+k_{n-4})}{q_{n-3} (k_1+...+k_i)},
\eeqn
where index $i$ denotes the factorization point between gluons with momenta $i$ and $i+1$.
Using the transverse momenta conservation
\beqn
q_1=k_1+...+k_{n-4}-q_{n-3}
\eeqn
we can  calculate
\beqn\label{wiplus1}
1+w_i=\frac{q_{n-3} (k_1+...+k_{i})+q_1 (k_{i+1}+...+k_{n-4})}{q_{n-3} (k_1+...+k_i)}.
\eeqn
The numerator of eq.~(\ref{wiplus1}) is simplified as follows
\beqn
&& q_{n-3} (k_1+...+k_{i})+q_1 (k_{i+1}+...+k_{n-4})=
(q_{i+1}-k_{i+1}-...-k_{n-4})(k_1+...+k_i) \nonumber
\\
&& +(q_{i+1}+k_1+...+k_i)(k_{i+1}+...+k_{n-4})=q_{i+1} (k_1+..+k_{n-4})
\eeqn
and we get
\beqn
1+w_i=\frac{q_{i+1} (k_1+..+k_{n-4})}{q_{n-3} (k_1+...+k_i)}.
\eeqn
We  also need
\beqn
1+\frac{1}{w_i}=\frac{q_{i+1} (k_1+..+k_{n-4})}{q_1 (k_{i+1}+...+k_{n-4})},
\eeqn
and finally write the corresponding $n=6$ remainder function in terms of the transverse momenta
\beqn
f_6 (w_i,w^*_i)=
\ln|1+w_i|^2 \ln \left|1+\frac{1}{w_i}\right|^2
=
\ln \frac{\mathbf{q}^2_{i+1}(\mathbf{k}_1+..+\mathbf{k}_{n-4})^2}{\mathbf{q}^2_{n-3} (\mathbf{k}_1+...+\mathbf{k}_i)^2}
\ln \frac{\mathbf{q}^2_{i+1} (\mathbf{k}_1+..+\mathbf{k}_{n-4})^2}{\mathbf{q}^2_{1} (\mathbf{k}_{i+1}+...+\mathbf{k}_{n-4})^2}.  \;\;\;\;
\eeqn
This result is valid only for the   $2 \to 2+(n-4) $   amplitude in the Mandelstam region, where we flip all $n-4$ produced gluons.

\end{document}